\documentclass[12pt]{iopart}
\usepackage{graphicx}
\usepackage{color}

\begin{document}

\title[Measuring stellar compactness with neutrinos]{Estimating the core compactness of massive stars with Galactic supernova neutrinos}

\author{Shunsaku Horiuchi$^1$, Ko Nakamura$^2$, Tomoya Takiwaki$^3$, Kei Kotake$^2$}
\address{$^1$Center for Neutrino Physics, Department of Physics, Virginia Tech, Blacksburg, VA 24061, USA}
\address{$^2$Department of Applied Physics, Fukuoka University, Fukuoka 814-0180, Japan}
\address{$^3$National Astronomical Observatory of Japan, Mitaka, Tokyo 181-8588, Japan}
\ead{horiuchi@vt.edu}
\vspace{10pt}
\begin{indented}
\item[]August 2017
\end{indented}

\begin{abstract}
We suggest the future detection of neutrinos from a Galactic core-collapse supernova can be used to infer the progenitor's inner mass density structure. We present the results from 20 axisymmetric core-collapse supernova simulations performed with progenitors spanning initial masses in the range $11$--$30M_\odot$, and focus on their connections to the progenitor compactness. The compactness is a measure of the mass density profile of the progenitor core and recent investigations have suggested its salient connections to the outcomes of core collapse. Our simulations confirm a correlation between the neutrinos emitted during the accretion phase and the progenitor's compactness, and that the ratio of observed neutrino events during the first hundreds of milliseconds provides a promising handle on the progenitor's inner structure. Neutrino flavor mixing during the accretion phase remains a large source of uncertainty.
\end{abstract}

%
\vspace{2pc}

\section{Introduction}\label{sec:intro}

Elucidating the mappings between core-collapse supernovae, their progenitor stars, and their compact remnants constitute a major goal of stellar and supernova research. The type of supernova an exploding star becomes depends strongly on the progenitor envelope and surrounding interstellar medium, yielding a wealth of rich supernova phenomenologies (e.g., \cite{Filippenko:1997ub}). On the other hand, it is increasingly evident that whether a star explodes or not, and what remnant it leaves behind, is strongly impacted by the progenitor's core structure \cite{Kitaura:2005bt,Kotake:2005zn,Janka:2006fh,O'Connor:2010tk,Burrows:2012ew}. 

Systematic simulations of the evolution of massive stars reveal large star-to-star variations in stellar interior structure \cite{Woosley:2002zz}. Quantities such as the iron core mass, mass density profile, entropy profile, and others appear to vary non-monotonic in zero-age main sequence (ZAMS) mass, and variations can be considerable even between stars separated by small mass differences \cite{Sukhbold:2013yca}. Recently, simulations of core collapse based on large numbers of progenitor initial conditions have been performed. One-dimensional spherically symmetric studies have paved the way, demonstrating that in the neutrino-driven delayed explosion mechanism, quantities such as the explosion energy, synthesized nickel mass, and remnant mass, as well as the outcome of core collapse into either neutron stars or black holes, also change non-monotonically with ZAMS mass, and that instead they can be more reliably predicted based on the so-called ``compactness parameter,'' which captures the density profile surrounding the collapsing core \cite{O'Connor:2010tk,Ugliano:2012kq,Pejcha:2014wda,Sukhbold:2015wba,Ertl:2015rga}. These trends have since been observed also in systematic two-dimensional axisymmetric simulations \cite{Nakamura:2014caa,Horiuchi:2014ska,Summa:2015nyk}. 

We define the compactness following O'Connor \& Ott \cite{O'Connor:2010tk} as, 
\begin{equation} \label{eq:compactness}
\xi_M = \left. \frac{M/M_\odot}{R(M_{\rm bary}=M)/1000\,{\rm km}} \right\vert_{ t },
\end{equation}
where $R(M_{\rm bary}=M)$ is the radial coordinate that encloses a baryonic mass $M$ at epoch $t$. The relevant mass scale in core collapse is $M=1.5$--$2.5 M_\odot$. The epoch of core bounce ($t=$ bounce) has been used by Ref.~\cite{O'Connor:2010tk}, but Sukhbold \& Woosley \cite{Sukhbold:2013yca} have shown that the pre-collapse progenitor compactness ($t=$ pre-collapse) works just as well for the values of $M$ of interest in core collapse, and we will adopt the latter's definition here. 

In the neutrino mechanism, a successful explosion occurs when the neutrino heating increases the pressure of the region below the stalled shock above the ram pressure of mass accretion impeding shock revival \cite{Kotake:2005zn,Janka:2006fh,Burrows:2012ew}. Given this balance, the concept of a critical curve has been useful in diagnosing the onset of neutrino-driven explosions \cite{Burrows:1993pi,Yamasaki:2006uz}. Namely, for a given mass accretion rate, there is a critical neutrino heating required for explosion; below this the shock cannot be revived. The critical neutrino heating as a function of mass accretion sets the critical curve, and importantly the curve depends on details of the microphysics, simulation setup, and explosion mechanism (e.g., see Ref.~\cite{Couch:2012kp} and references therein). Note, the core-collapse process is a dynamical phenomenon, and the concept of a critical curve works as well as it does provided one focuses on the critical time epoch of shock revival. 

The compactness is useful since by choosing an appropriate mass $M$, it is able to characterize the mass accretion during the epochs of shock revival. The relevant mass scale is between 1.5--$2.5 M_\odot$ \cite{O'Connor:2010tk,Sukhbold:2013yca}. Physically, a progenitor with larger compactness  will have a more compact higher density core, which results in a longer-lasting higher mass accretion rate, working to suppress shock revival. Thus, progenitors with large compactness tend to be more difficult to explode and more prone to collapsing to black holes \cite{O'Connor:2010tk}. The precise compactness value where the transition occurs---which we term the \textit{critical} compactness---depends on the explosion mechanism as well as the simulation setup. For the neutrino mechanism, systematic simulations performed under spherically symmetric and axisymmetric geometries show it to fall in the range $\xi_{\rm 2.5,crit} \approx 0.2$--0.5 for $M=2.5 M_\odot$ and a single critical value succeeds in predicting the core collapse outcome with a 80--90\% success rate \cite{O'Connor:2010tk,Ugliano:2012kq,Horiuchi:2014ska,Pejcha:2014wda,Sukhbold:2015wba,Ertl:2015rga}. The large range is in part due to the need for modeling the fact that spherically symmetric simulations require artificial heating to explode, and axisymmetric simulations are too conducive to explosions by reverse cascading of perturbation power. In the future, suites of simulations in full three-dimensional geometry and improved microphysics will provide better insights of whether one can define a useful critical compactness, and if so, what its value is. 

It should be warned that there are limitations to diagnosing the complex core-collapse process using a single parameter. For example, the temporal history of mass accretion is important. This is because broadly speaking, the neutrino emission from the central neutrinosphere is set by the past history of mass accretion which dictates the amount of gravitational energy available, while the ram pressure the shock must overcome is set by the ongoing mass accretion through the stalled shock. For example, a progenitor with a rapidly declining mass accretion history is conducive to explosions \cite{Kitaura:2005bt,Huedepohl:2009wh,Radice:2017ykv}, although due to the smaller total mass accretion the explosion energetics tend to be small. Thus in general, two parameters are required for a fuller description---one capturing the protoneutron star mass and the other capturing the mass accretion rate during the critical time of supernova shock revival---and Ertl \textit{et al.}~\cite{Ertl:2015rga} have shown that indeed two parameters improve the predictive success rate to as high as 97\%. 

Nevertheless, the simplicity of the compactness has its merits. It has already been connected to the discussion of several astronomical observations. All are concerned with the outcome of core collapse as either supernova explosions leaving behind neutron stars or optically dark or dim ``failed explosions'' leaving behind black holes. For example, Smartt and collaborators \cite{Smartt:2008zd,Smartt:2015sfa} have quantified the dearth of the most massive stars with ZAMS mass above $16.5 M_\odot$ discovered as progenitors of Type IIP supernovae. Given that red supergiants with higher estimated mass exist in the Local Group, it begs the question what happens to these most massive giants? One possibility is that they undergo failed explosions (other possibilities exist, e.g., Refs.~\cite{Walmswell:2011us,Yoon:2012yk,Groh:2013mma}). Interestingly, stellar evolution theory predicts that stars with mass around $\sim 20 M_\odot$ inhabit a peak in compactness \cite{Sukhbold:2013yca}, which would be consistent with them tending to fail \cite{Horiuchi:2014ska,Kochanek:2014mwa}. Quantitatively, the critical compactness would need to be low, around $\xi_{\rm 2.5,crit} \sim 0.2$ \cite{Horiuchi:2014ska}. Secondly, an ongoing survey is currently searching for failed explosions more directly by locating the quiet disappearance of massive stars, i.e., without a supernova \cite{Kochanek:2008mp}. In 7 years of operation, the survey has discovered one candidate \cite{Gerke:2014ooa}, which yields a fraction of failed supernovae of 4--43\% at 90\%CL \cite{Adams:2016ffj}, a value that is consistent with $\xi_{\rm 2.5,crit} = 0.2$ (see Figure 1 bottom panel of Ref.~\cite{Horiuchi:2014ska}). Another observational indication is the apparent deficit of the observed core-collapse supernova rate when compared to the birth rate of massive stars \cite{Horiuchi:2011zz}. The fraction of missing massive stars is consistent with $\xi_{\rm 2.5,crit} = 0.2$ \cite{Horiuchi:2014ska}, although uncertainties are still significant \cite{Kobayashi:2012es,Mathews:2014qba}. Finally, a similarly low critical compactness can also explain the mass function of neutron stars and black holes \cite{Kochanek:2013yca,Kochanek:2014mwa}. While individually amounting to a hint at best, it is intriguing that these results together paint a consistent picture in terms of critical compactness. It is therefore of interest to consider ways of more directly probing the compactness---and more generally speaking, the core properties---of core-collapse progenitors.

Unfortunately, the core of a supernova progenitor is hidden by many solar masses of opaque stellar material, and core properties are largely decoupled from the envelope, making electromagnetic probes difficult. In this article, we demonstrate that the neutrino emission during the core collapse is itself an indicator of the progenitor compactness, and we propose a strategy to reveal the progenitor compactness using neutrino observations of a future Galactic core collapse.

In \Sref{sec:setup}, we introduce our simulation setup and implementation of neutrino flavor mixing. In \Sref{sec:ratio}, we present neutrino event rate predictions and show a simple way how they can be used to probe the progenitor core compactness. We also discuss sources of uncertainties. In \Sref{sec:discussion}, we summarize and conclude.

\section{Setup}\label{sec:setup}

\subsection{Numerical simulation}\label{sec:signal}

We take the two-dimensional axisymmetric core-collapse models from Nakamura \textit{et al.}~\cite{Nakamura:2014caa} with several upgrades. In these models, self-gravity was computed by a Newtonian monopole approximation, now upgraded with effective General Relativistic corrections. Neutrino transport is also upgraded with respect to Ref.~\cite{Nakamura:2014caa}: it is solved with an energy-dependent treatment of neutrino transport based on the isotropic diffusion source approximation (IDSA, \cite{Liebendoerfer:2007dz}) with a ray-by-ray approach for all neutrino species ($\nu_e$, $\bar{\nu}_e$, and $\nu_x$, where $\nu_x$ refers to heavy-lepton flavor neutrinos and anti-neutrinos). This approximation has a high computational efficiency in parallelization, which allows to explore systematic features of neutrino emission for a large number of supernova models whilst maintaining high accuracy results. The equation of state (EOS) by Lattimer \& Swesty \cite{Lattimer:1991nc} with incompressibility of 220 MeV is adopted. 

In Ref.~\cite{Nakamura:2014caa}, 378 non-rotating progenitor stars from Woosley \textit{et al.}~\cite{Woosley:2002zz} covering ZAMS mass from 10.8 $M_\odot$ to 75 $M_\odot$ with metallicity from zero to solar value were investigated. From these, we choose 20 progenitor models with solar metallicity for the current study, with masses $11 M_\odot$ to $30 M_\odot$ in steps of $1M_\odot$. We neglect lower metallicity progenitors given that they are rare in the local Universe. The chosen 20 models cover a wide range of compactness, $\xi_{2.5}$ from 0.0039 for the $11 M_\odot$ model to 0.434 for the $23 M_\odot$ model.

\subsection{Signal calculation}\label{sec:signal}

From the simulations we extract three neutrino emission parameters---luminosity $L_\nu$, mean energy $\langle E_\nu \rangle$, and spectral pinching parameter $\alpha$---for three neutrino species $\nu_e$, $\bar{\nu}_e$, and $\nu_x$. We define the pinching parameter by the first and second moments of the neutrino energy spectrum,
\begin{eqnarray} \label{eq:}
\alpha(t)  &=& \frac{2\langle E_{\nu}\rangle^2 - \langle E_{\nu}^2 \rangle} {\langle E_{\nu}^2 \rangle - \langle E_{\nu} \rangle^2} ~,
\end{eqnarray}
where $\langle E_\nu \rangle$ is the mean energy and $\langle E_{\nu}^2\rangle$ is the mean square energy. These parameters provide an accurate analytic description of the neutrino spectrum \cite{Keil:2002in,Tamborra:2012ac},
\begin{eqnarray}\label{eq:pinchedFD}
f(E_\nu)= \frac{(1+  \alpha )^{(1+  \alpha )}}{\Gamma(1+  \alpha )} \frac{ E_\nu^{ \alpha }}{\langle E_\nu \rangle^{1 + \alpha }} {\rm exp}\left[ {-(1+  \alpha ) \frac{E_\nu}{ \langle E_\nu \rangle }} \right].
\end{eqnarray}
The flux of neutrinos from a Galactic supernova is then,
\begin{eqnarray}\label{eq:pinchedFD}
\frac{dF_\nu}{dE_\nu}(E_\nu)= \frac{1}{4 \pi d^2} \frac{L_\nu}{\langle E_\nu \rangle} f(E_\nu),
\end{eqnarray}
where $d=10$ kpc is the distance to the supernova.

The neutrinos that are emitted from the neutrinospheres undergo flavor mixing during their propagation to a terrestrial detector. The most well-understood are vacuum oscillations and the matter-induced MSW effect, which result in a $\nu_e$ survival probability of 0 and a $\bar{\nu}_e$ survival probability of $\cos^2 \theta_{12}$, both for the normal mass hierarchy \cite{Dighe:1999bi,Mirizzi:2015eza}. Here, $\theta_{12}$ is the solar mixing angle and $\sin^2 \theta_{12} \simeq 0.3$. Thus, the terrestrial fluxes of $\nu_e$ and $\bar{\nu}_e$ are,
\begin{eqnarray}
F_{\nu_e}& \simeq &F^0_{\nu_x} \label{eq:MSWnorm} \\
F_{\bar{\nu}_e}& \simeq& \cos^2 \theta_{12} F^0_{\bar{\nu}_e} + \sin^2 \theta_{12} F^0_{{\nu}_x},
\end{eqnarray}
where we explicitly denoted by the superscript $0$ the fluxes emitted from the neutrinospheres, and have omitted the $\theta_{13}$ part since $\sin^2 \theta_{13} \approx 0.02$. We also omit Earth effects for simplicity and generality. For inverted mass hierarchy the survival probability change and the fluxes become,
\begin{eqnarray} 
F_{\nu_e}& \simeq &\sin^2 \theta_{12} F^0_{\nu_e} + \cos^2 \theta_{12} F^0_{\nu_x}  \\
F_{\bar{\nu}_e} &\simeq &F^0_{\bar{\nu}_x}.\label{eq:MSWinv}
\end{eqnarray}

The situation is however more complicated, as additional flavor mixing can be induced by the coherent neutrino-neutrino forward scattering potential. The accretion phase, when the flux ordering is $F^0_{\nu_e} > F^0_{\bar{\nu}_e} > F^0_{\nu_x}$, is conducive to this so-called collective oscillations. However, the precise predictions of flavor mixing, and its dependence on neutrino energy, is far from certain (for reviews, see, e.g., Refs.~\cite{Duan:2010bg,Mirizzi:2015eza}). For example, Ref.~\cite{Dasgupta:2007ws} explored the single-angle approximation in a 3 flavor framework, and showed that collective oscillations operate in the accretion phase only in the inverted mass hierarchy with $\nu_e$ and $\bar{\nu}_e$ survival probabilities of 0 and $\cos^2 \theta_{12}$, both above a critical energy of $\sim 10$ MeV. On the other hand, Ref.~\cite{Mirizzi:2010uz} extended this to a multi-angle treatment and showed survival probabilities of 0 and 1 above critical energies of $\sim 10$ MeV and $\sim 3$ MeV, respectively. Furthermore, additional effects such as the halo effect \cite{Cherry:2012zw} and fast flavor conversions \cite{Sawyer:2005jk} may be potentially important, and a complete picture of the self-induced flavor conversion is still missing. 

To remain open to novel mixing phenomena, we follow Dasgupta \textit{et al.}~\cite{Dasgupta:2009yj} and adopt three effective ``MSW + collective mixing'' scenarios for the accretion phase: ($i$) $\bar{\nu}_e$ survival probability of 0, or full swap, ($ii$) $\bar{\nu}_e$ survival probability of 1, or no mixing, and ($iii$) $\bar{\nu}_e$ survival probability of $\cos^2 \theta_{12}$. While adopting no to full mixing may be extreme, it works to fill the range of predictions possible. Note that the MSW-only phenomenology is included as a subset in our three scenarios: the normal mass hierarchy scenario ($iii$), and the inverted mass hierarchy scenario ($i$). We adopt the same approach for $\nu_e$, which leads to us adopting survival probabilities of 0, 1, and $\sin^2 \theta_{12}$. 

Self-induced effects do not operate during the neutronization burst due to the large excess of $\nu_e$ flux \cite{Hannestad:2006nj}. We therefore only implement MSW mixing using Eqs.~(\ref{eq:MSWnorm})--(\ref{eq:MSWinv}) for the early phase. We define the transition from the neutronization to accretion phase by when the flux of $\bar{\nu}_e$ overtakes the $\nu_x$ flux, which typically occurs within the first $\sim 20$ ms post-bounce. 

For detection, we first consider water Cherenkov detectors Super-Kamiokande (Super-K) and Hyper-Kamiokande (Hyper-K), with 32 kton and 440 kton inner detectors and lepton detection thresholds of 3 MeV and 5 MeV, respectively. In both cases, perfect efficiency above the detection threshold is assumed. We focus on the inverse-$\beta$ decay (IBD) process which dominates the event statistics in these detectors. Later, we will consider other detectors which provide complementary neutrino flavor information. 

\begin{figure}
\includegraphics[width=120mm,bb=0 50 800 590]{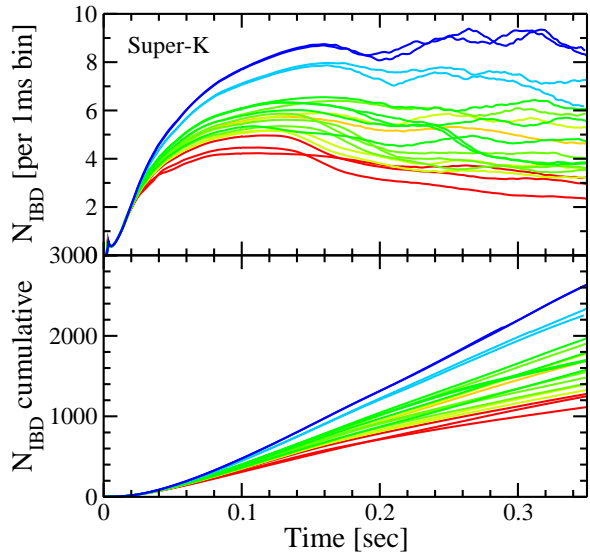}
\caption{IBD event rate predictions for Super-K, shown per 1 msec bins (top panel) and cumulative (bottom panel). Each curve is an axisymmetric simulation with color coding corresponding to progenitor compactness, from $\xi_{2.5} = 0.0039$ (red) to 0.434 (blue). MSW mixing under normal mass hierarchy is adopted.}
\label{fig:Nrate}
\end{figure}

\Fref{fig:Nrate} shows the IBD events at Super-K for a Galactic core collapse at a distance of 10 kpc. The top panel shows the events in 1 msec time bins, and the bottom panel shows the cumulative events. Each curve represents a 2D simulations, color coded by the progenitor compactness $\xi_{2.5}$, ranging from 0.0039 (red) to 0.434 (blue). A trend with compactness is observed, with progenitors with larger compactness yielding higher neutrino events. The reason is because higher compactness leads to higher mass accretion rate, which leads to more gravitational energy liberation and hence higher neutrino energetics. This correlation is consistent with what is seen in spherically symmetric simulations as shown by O'Connor \& Ott \cite{OConnor:2012bsj}. That the trend holds also in axisymmetric simulations, where the additional degree of spatial freedom means asymmetric mass accretion is possible (and indeed frequently seen, e.g., \cite{Nakamura:2014caa}), shows the strong impact of the progenitor mass density profile on the neutrino emission. For illustrative purposes, only the normal mass hierarchy with MSW oscillation scenario is shown in \Fref{fig:Nrate}, but the trends remain in other scenarios.

\section{Estimating the compactness}\label{sec:ratio}

\Fref{fig:Nrate} shows that a detailed measurement of the neutrino event rates could reveal the progenitor compactness. Given that Super-K is expected to detect $O(10^4)$ IBD events from a Galactic core collapse, the number of events in principle will be measured with percent precision. However, the comparison with models is subject to many other systematic uncertainties, including detector efficiency, oscillation uncertainties, and distance uncertainties. For example, distance uncertainties are often at least several tens of percent (see, e.g., Ref.~\cite{Rosslowe:2005aa} for stripped massive stars and Ref.~\cite{Nakamura:2016kkl} for supergiant stars). Even with electromagnetic observations of a supernova, measures of distances are not easy, e.g., by observing the expanding remnant \cite{Allen:2014yra}. Without any electromagnetic observations, the distance measurement must rely on other means, e.g., gravitational wave detection or the neutrino emission itself. Using the neutrino signal to estimate both the compactness \textit{and} the distance may appear to be a circular argument. Fortunately, the initial phase of the core collapse provides a phase of the neutrino emission that can be used (close to) a standard candle, more or less independently of the accretion phase. Since the collapse is still close to spherical, the early phase neutrino emission depends weakly on the progenitor structure as well as setup of numerical calculation (e.g., dimensionality). For example, the neutronization bust is emitted during the first $\sim 20$ msec post-bounce as an intense pulse of $\nu_e$ reaching luminosities of several $10^{53} \, {\rm erg/s}$, when the bounce shock propagates across the neutrinosphere and rapidly reduces the neutrino opacity by dissociating surrounding nuclei. Kachelriess \textit{et al.}~\cite{Kachelriess:2004ds} showed that the neutronization burst can be used to determine the distance to a supernova with an accuracy of $\sim 5$\% by next-generation Mton-class detectors. 

\begin{figure}
\includegraphics[width=120mm,bb=0 50 800 590]{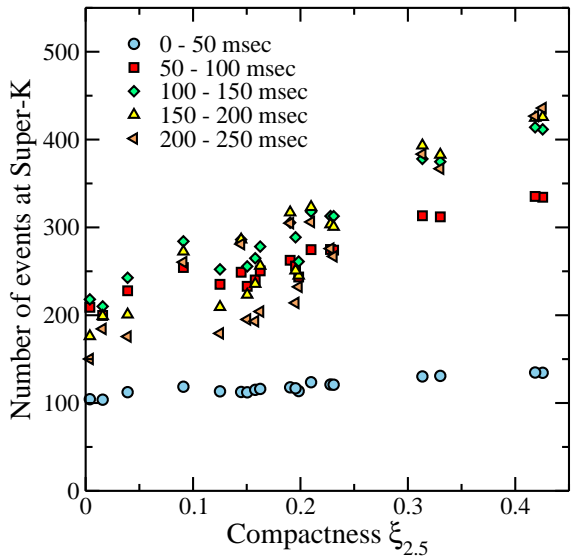}
\caption{The number of IBD events predicted in Super-K in five time windows of 50 msec duration, as functions of progenitor compactness $\xi_{2.5}$. The earliest time window shows weak dependence on the compactness, while later epochs show strong dependence due to the progenitor-dependent mass accretion rates.}
\label{fig:Ntotal}
\end{figure}

\Fref{fig:Ntotal} shows the predicted number of IBD events in several different 50 msec time windows all as functions of the progenitor compactness. Once again the Super-K and a MSW oscillation scenario under normal mass hierarchy is adopted. It is evident that the number of events during the first 50 msec is close to independent of the compactness $\xi_{2.5}$, while later time windows show the expected rise with compactness. 

\begin{figure}
\includegraphics[width=160mm,bb=0 50 800 590]{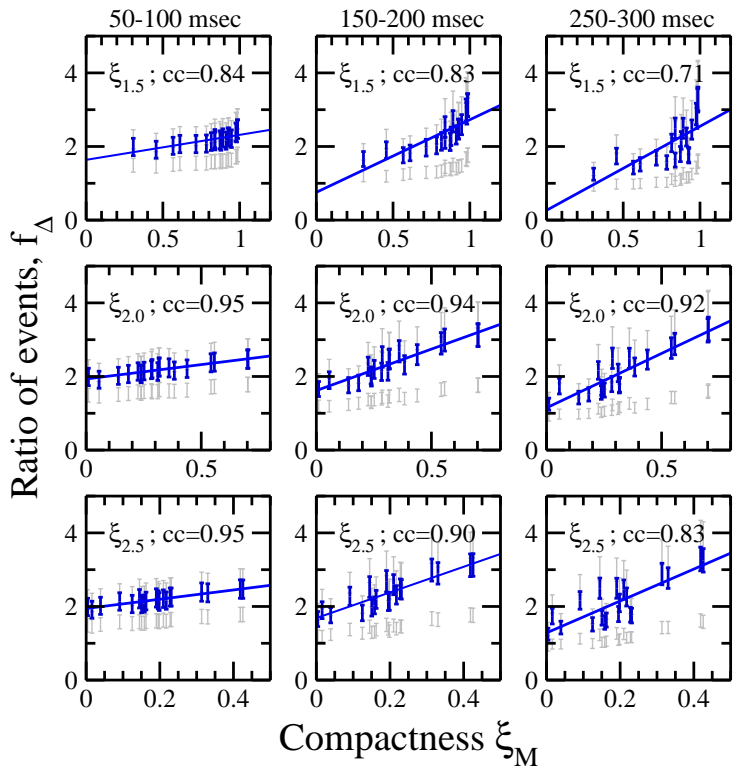}
\caption{Predicted ratio of IBD events at Super-K, calculated for multiple time windows: the left, center, and right columns show the 50--100 msec, 150--200 msec, and 250--300 msec windows, respectively, all divided by the number of events in the 0--50 msec window. The ratios are plotted as functions of progenitor compactness: the top, central, and bottom rows show $\xi_{1.5}$, $\xi_{2.0}$, and $\xi_{2.5}$, respectively. Error bars are statistical errors. Blue points show MSW mixing. Straight lines are linear fits to the blue points, with the correlation coefficient labeled for reference. In gray, we show predictions adopting an extreme $\bar{\nu}_e$ survival probability of 0 and 1 during the accretion phase, for illustration.}
\label{fig:norm3by3}
\end{figure}

Based on this result, we show in \Fref{fig:norm3by3} the ratio of the total events in a time window $\Delta t$ over the total events in the first 50 msec, 
\begin{equation}\label{eq:ratio}
f_\Delta = \frac{N(\Delta t)}{N(0-50\,{\rm msec})},
\end{equation}
and plot them against compactness defined by three values of mass $M= 1.5 M_\odot$, $2.0 M_\odot$, and $2.5 M_\odot$, shown on each row. Each column corresponds to a different time window $\Delta t$ of 50--100 msec, 150--200 msec, and 250--300 msec. The blue points represent MSW mixing under normal mass hierarchy. The error bars shown are statistical errors only, and are dominated by the denominator $N(0-50\,{\rm msec})$ due to its generally smaller total number of events compared to later time windows. Since the distance uncertainty---and indeed any systematic uncertainty that affects both $N(\Delta t)$ and $N(0-50\,{\rm msec})$ equally---cancel, the $y$-axis is a measurable quantity that can more robustly be compared with predictions to infer the compactness. 

In general the ratios correlate strongly with compactness. In \Fref{fig:norm3by3} we label each panel by the correlation coefficient, defined
\begin{equation}\label{eq:cc}
cc = \frac{\sum_i (\xi_i - \bar{\xi})(f_{\Delta,i} - \bar{f}_\Delta)   }
{\sqrt{\sum_i (\xi - \bar{\xi})^2} \sqrt{\sum_i (f_{\Delta,i} - \bar{f_\Delta})^2} },
\end{equation}
where bars indicate arithmetic means. The correlation coefficients are generally high. In terms of the response of the correlation (i.e., the slope), later time windows are generally better indicators of compactness defined by larger $M$. This can be understood by the fact that the neutrino emission in later epochs is associated with the accretion of mass shells of higher mass coordinate, at least until shock revival (after shock revival, mass accretion is largely halted). In other words, we expect each time window to hold information about a particular range of mass coordinates, which is to say, it holds information about a particular range of compactness definitions, with later epochs probing larger $M$. A one-to-one mapping is beyond the scope of this paper as it would require considering the accretion time of a mass element $dM$ at radius $R$ to the protoneutron star, together with the time-delay in converting the gravitational binding energy liberated to neutrino emission. Given the angular dependence of mass accretion in multi-dimensional simulations, average values would  need to be defined that picks out the the bulk of the mass accretion and conversion to neutrino emission. Nevertheless, the appearance of correlations in \Fref{fig:norm3by3} encouragingly shows that despite the possibilities of large asphericities, there are correlations between the progenitor mass density structure and the neutrino light curve. Thus, we can observe how the compactness may be inferred from the detected neutrinos. For example, the compactness $\xi_{2.0}$ can be inferred best from intermediate epochs, e.g., the 150--200 msec time window (central panel). 

There are a number of potential sources of uncertainties that would complicated the interpretation of measured neutrino event ratios. In \Fref{fig:norm3by3}, the blue points show the results under MSW mixing. However, as discussed in \Sref{sec:signal}, additional oscillation features are possible, which can potentially have a significant impact on the predicted ratios. The gray points in \Fref{fig:norm3by3} show predictions based on rather extreme assumptions of no mixing and full mixing during the accretion phase. Although the range of realistic possibilities may be smaller than the range of ratios shown, it is evident that improved understanding of neutrino mixing effects during the accretion phase would be important to improve delineating the compactness from future neutrino datasets. Other potential uncertainties, not directly addressed in this paper, include the effects of stellar core rotation, the impact of full three-dimensional modeling, and updated treatments of the EOS of dense hot matter. Depending on which epochs these effects impact the neutrino emission, they can alter predictions of neutrino event ratios. 

\begin{figure}[t]
\includegraphics[width=160mm,bb=0 350 800 600]{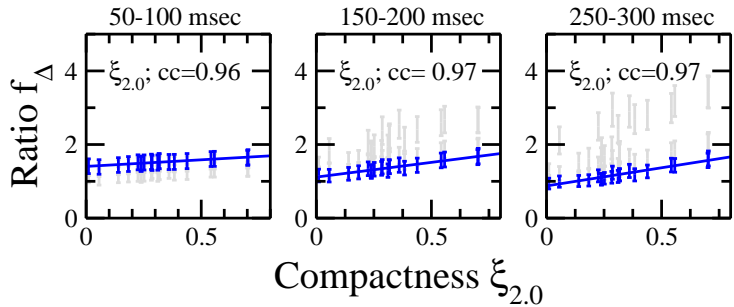}
\caption{The same as the central row of \Fref{fig:norm3by3}, but for $\nu_e$ detection by DUNE via CC interaction on Ar.}
\label{fig:norm3by1}
\end{figure}

In Figure \ref{fig:norm3by1}, we show the event ratios Eq.~(\ref{eq:ratio}) for $\nu_e$ as a probe of the compactness $\xi_{2.0}$ (i.e., the analogous of the central row of \Fref{fig:norm3by3}), focusing on the charged-current interaction $\nu_e + {}^{40}{\rm Ar} \to e^- + {}^{40}{\rm K}^*$ relevant for Deep Underground Neutrino Experiment (DUNE). We assume a fiducial volume of 40 kton, a detection threshold energy of 5 MeV, and that the events will be isolated via a photon system targeting the decay photons of ${}^{40}{\rm K}^*$. The left, central, and right panels show the time windows 50--100 msec, 150--200 msec, and 250--300 msec, respectively. Once again, the blue points show predictions based on MSW mixing, while the gray points show extremes of beyond-MSW mixing during the accretion phase. Each panel is labeled by the correlation coefficient for the MSW mixing predictions, using Eq.~(\ref{eq:cc}). The predicted range of ratios is large, due to the fact the $\nu_x$ with which the $\nu_e$ mix has significantly higher mean energies than the $\nu_e$, implying mixing uncertainties translate to large differences to event rate predictions. Once mixing uncertainties are under better control, the ratios of $\nu_e$ events at DUNE will prove a good measure of the progenitor compactness. One strength of $\nu_e$ is in the fact that it contains the neutronization burst, which allows one to define neutrino ratios based on a better-defined time window using the neutronization burst. 

\section{Discussions and conclusions}\label{sec:discussion}

Neutrinos offer a unique and powerful way to view the interiors of stars. We have presented a simple way of using neutrinos to probe the core compactness of massive stars undergoing core collapse. Even a simple ratio of neutrino event rates will be useful for revealing whether the progenitor undergoing core collapse has a large compactness or not. The inferred value of the compactness will in turn be useful to test core-collapse models. Recent theoretical investigations have shown that the compactness is a simple yet useful parameter to discuss the outcomes of core collapse, and multiple studies have suggested that there may be a critical compactness---albeit with large uncertainty still---beyond which massive stars fail to explode and instead collapse to black holes \cite{O'Connor:2010tk,Ugliano:2012kq,Nakamura:2014caa,Horiuchi:2014ska,Pejcha:2014wda,Summa:2015nyk,Sukhbold:2015wba,Ertl:2015rga}. The progenitor compactness inferred from neutrino, coupled with the observation (or null observation) of a supernova, can test such scenarios. 

There remain many uncertainties that must be addressed in the future. We have explored the impact of additional neutrino flavor mixing beyond MSW. Future investigations of mixing effects during the accretion epoch will be important to reduce the uncertainty in neutrino event predictions. Also, our results were based on axisymmetric hydrodynamic simulations adopting a single equation of state. However, there is still some degree of uncertainty in the nuclear physics that is relevant for the early phases of core collapse. For example, the peak of the neutronization burst can vary by some $\pm 20$\% when progenitor and EOS variations are considered \cite{Sullivan:2015kva}. Predictions will need to be constantly updated with improved microphysics and also eventually explored with full three-dimensional simulations when they become feasible to study multiple progenitor initial conditions. Such improvements will enable better predictions that can be used to do what neutrinos do best---infer the properties of stellar interiors. 

\section*{Acknowledgments}

This study was supported by JSPS (Nos. 24103006, 24244036, 26707013 and 26870823) and by MEXT (Nos. 15H00789, 15H01039, 15KK0173, 17H01130, 17H06364, 17H06357) and JICFuS as a priority issue to be tackled by using Post ‘K’ Computer. 

\bibliographystyle{unsrt}
\bibliography{ms_compactness_iop_resubmit.bbl} 


\end{document}